\documentclass[10pt,epsf]{article}

\usepackage{graphicx}
\usepackage{indentfirst}

\begin{document}

\title{\bf{Particle Probe of Ho\v{r}ava-Lifshitz Gravity}}
\date{}
\maketitle

\begin{center}\author{Bogeun Gwak}\footnote{rasenis@sogang.ac.kr} and \author{Bum-Hoon Lee}\footnote{bhl@sogang.ac.kr}\\ \vskip 0.25in $^{1,2}$\it{Department of Physics and Center of Quantum Spacetime, Sogang University, Seoul 121-742, Korea} \end{center} \vskip 0.6in

{\abstract
{Kehagias-Sfetsos black hole in Ho\v{r}ava-Lifshitz gravity is probed through particle geodesics.  Gravitational force of KS black hole becomes weaker than that of Schwarzschild around horizon and interior space. Particles can be always scattered or trapped in new closed orbits, unlike those falling forever in Schwarzschild black. The properties of null and timelike geodesics are classified with values of coupling constants. The precession rates of the orbits are evaluated. The time trajectories are also classified under different values of coupling constants for both null and timelike geodesics. Physical phenomena that may be observable are discussed.}}
\thispagestyle{empty}
\newpage

\setcounter{page}{1}
\section{Introduction}

Recently, UV complete renormalizable theory of gravity at Lifshitz point is suggested by Ho\v{r}ava\,\cite{bib1}. The gravity theory named Ho\v{r}ava-Lifshitz\,(HL) theory has a nonrelativistic symmetry in UV region avoiding nonrenormalizability from Einstein gravity. It is claimed to recover to Einstein gravity in IR region in specific case. One may impose the detailed balance condition and projectability condition. Its coupling constants and symmetry is studied\,\cite{bib2d}. The theory is known to give various solutions\,\cite{bib3} and black hole thermodynamics\,\cite{bib2h}. It is expected that the black hole solutions in HL-gravity asymptotically become Einstein gravity solutions. A  black hole solution by A. Kehagias and K. Sfetsos\,(KS) is one of them\,\cite{Kehagias:2009is}. It becomes Schwarzschild black hole in asymptotic region. It can be regarded as a higher derivative modified solution of Schwarzschild black hole in Einstein gravity. Geodesics are stable in KS solution\,\cite{bib4}. The effective potential at fixed coupling constants is also studied\,\cite{bib4a}. Three GR tests are suggested in this solution with fixing coupling constants to match experimental observation of the bending of light\,\cite{bib4b}. It may be used as applications for solar system and constraints of HL-theory\,\cite{bib6}. There are various cosmology solutions\,\cite{bib2}, and its perturbations\,\cite{bib2a}. Also, dark matter and energy in HL-gravity are studied\,\cite{bib2g}. HL-gravity causes attentions to a lifshitz point\,\cite{bib2b}.

The motion of high energy particle in curved spacetime is important. The motions of probes is given by geodesics. Through the motions, the physical and geometrical quantities can be probed. The particle motion is well described in small curvature geometry by Einstein gravity. The higher derivative terms become important in the high curvature region. By probe this region, the deviation from Einstein gravity can be detectable. In this paper, particle motions in KS black hole of HL-gravity are studied to show the difference between HL-gravity and Einstein gravity. The proper approach for understanding spacetime structure related to particle behaviors is geodesics in KS black hole. The objects such as galactic nuclei and neutron stars are described using Einstein gravity. The difference of gravitation compared with Schwarzschild solution will modify physics of cosmological objects. From the higher derivative, the KS metric is dependent on the coupling constants which is written $\omega\,$. Since each coupling constants represents different spacetimes, the geodesic properties are classified by constant $\omega$. 

This paper is organized as follows. First, we introduce shortly the KS black hole. Secondly, the effective potentials and full set of geodesic equations for null and timelike case are obtained and evaluated with comparing to both KS and Schwarzschild black holes. The gravitational force is analyzed in and around horizons. The geodesic orbits are numerically classified with $\omega$ for null and timelike cases. There are new orbits observed in this classification. Following orbit classification, the time structures is categorized with $\omega$ showing the time structure suffered by particle falling to horizon. Finally, we summarize our results and discuss possible observation effects.

\section{KS black hole}
The 4-dimensional gravity action by Ho\v{r}ava-Lifshitz is given by,
\begin{eqnarray}
\label{action2}
S=&&\int dtdx^3\sqrt{g}N\Huge[\frac{2}{\kappa^2}(K_{ij}K^{ij}-\lambda K^2)-\frac{\kappa}{2\omega^4}C_{ij}C^{ij}+\frac{\kappa^2\mu}{2\omega^2}\epsilon^{ijk}R^{(3)}_{il}\bigtriangledown_{j}R^{(3)l}_{k}-\frac{\kappa^2 \mu^2}{8}R_{ij}^{(3)}R^{(3)ij}\\&&+\frac{\kappa^2 \mu^2}{8(1-3\lambda)}\left(\frac{1-4\lambda}{4}(R^{(3)})^2+\Lambda_W R^{(3)}-3\Lambda^2_W\right)+\mu^4R^{(3)}\Large]\,\nonumber.
\end{eqnarray}
The last term in eq.(\ref{action2}) breaks softly detailed balanced condition. It still satisfies the detailed balance condition in UV region. The action for $\Lambda_W=0\,$ gives solutions that are asymptotically flat metric. In addition, for $\lambda=1\,$, the black hole type solution by KS is written as\,\cite{Kehagias:2009is}, 
\begin{eqnarray}
&&ds^2=-f(r)dt^2+\frac{dr^2}{f(r)}+r^2(d\theta^2+\sin^2\theta d\phi^2)\, ,\\
&&f(r)=1+\omega r^2-\sqrt{r(\omega^2 r^3+4\omega M)}\,,
\end{eqnarray}
in which the $\omega=16\mu^2/\kappa^2$. This spacetime becomes a Schwarzschild black hole for $r\gg (M/\omega)^{\frac{1}{3}}$, i.e. for large $r$ or large $\omega$ in fixed $r$. The Kretschmann scalar, $R^{\mu\nu\rho\lambda}R_{\mu\nu\rho\lambda}$, diverges only at $r=0$, so the singularity is positioned at the center of spacetime. There exist two event horizons for $\omega M^2> \frac{1}{2}$ at
\begin{eqnarray}
r_{\pm}=M\left(1\pm\sqrt{1-\frac{1}{2\omega M^2}}\right)\,.
\end{eqnarray}
It has one horizon for $\omega M^2= \frac{1}{2}$. The singularity is naked for $\omega M^2< \frac{1}{2}$. In other words, the singularity is covered by horizons for $\omega M^2\geq \frac{1}{2}$. The effect of higher derivatives terms is controlled by the coupling $\omega\,$.

In our analysis, we rescale the parameters as follows,
\begin{eqnarray}
&&\tilde{r}=\frac{r}{2M}\,,\,\,\,\,\tilde{\omega}=\omega M^2\,.
\end{eqnarray}
In this new scaling, the function $f(r)$ and the positions of horizons are rewritten,
\begin{eqnarray}
\tilde{r}_{\pm}=\frac{1}{2}\left(1\pm\sqrt{1-\frac{1}{2\tilde{\omega}}}\right)\,,\,\,\,\,f(\tilde{r})=1+4\tilde{\omega}\tilde{r}^2 -\sqrt{16\tilde{\omega}^2\tilde{r}^4+8\tilde{\omega}\tilde{r}}\,.
\end{eqnarray}
Schwarzschild horizon is positioned at $\tilde{r}=1$ in asymptotic limit. For simplicity, tildes will be omitted. The horizons have $\omega$-dependency. Since Schwarzschild horizon is fixed at $r=1$ in this work, the outer horizon, $r_+$, which is always smaller than Schwarzschild case becomes to $r=1$ in large limit of $\omega$ or $r$. In the same limit, the inner horizon, $r_-$, is getting closer to the singularity at $r=0$. On contrary, higher curvature terms depending on $\omega$ modifies geometry around horizon and its interior structure. In this context, comparing geometries of KS and Schwarzschild black hole shows how gravitation is changed  with the scale.

\section{Spacetime Geometry}
\begin{figure}[h]
\centering
\includegraphics[scale=0.5,keepaspectratio]{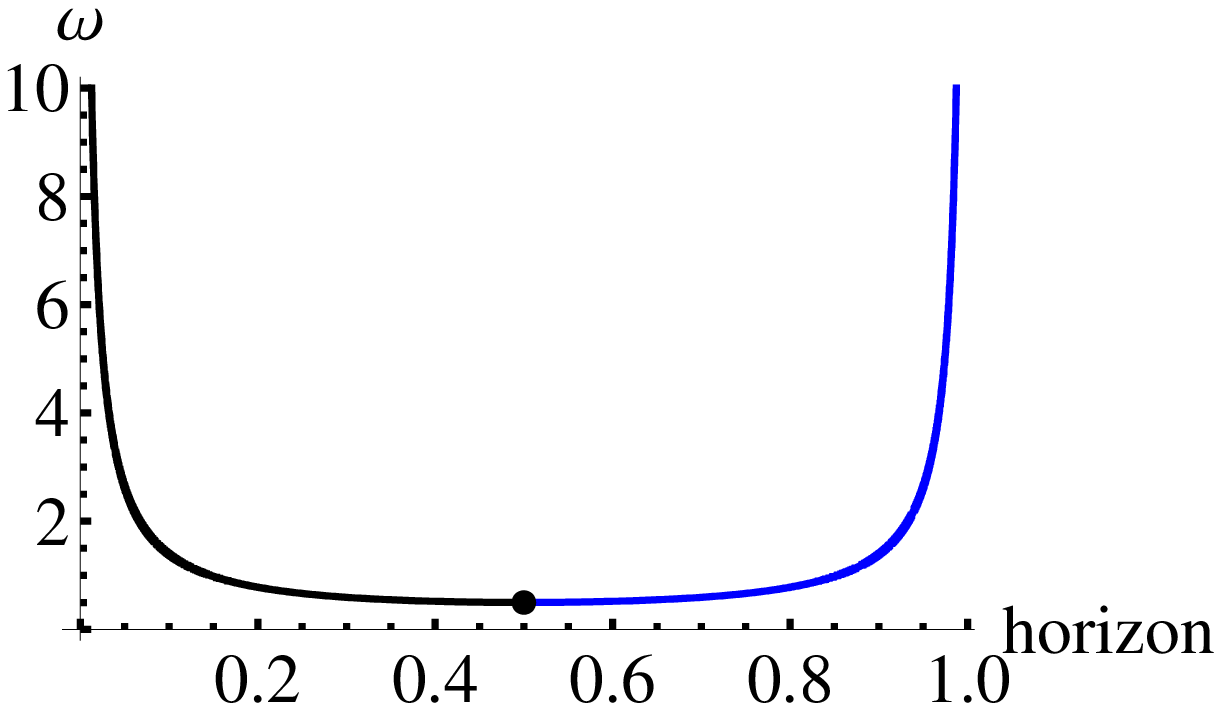}
\includegraphics[scale=0.5,keepaspectratio]{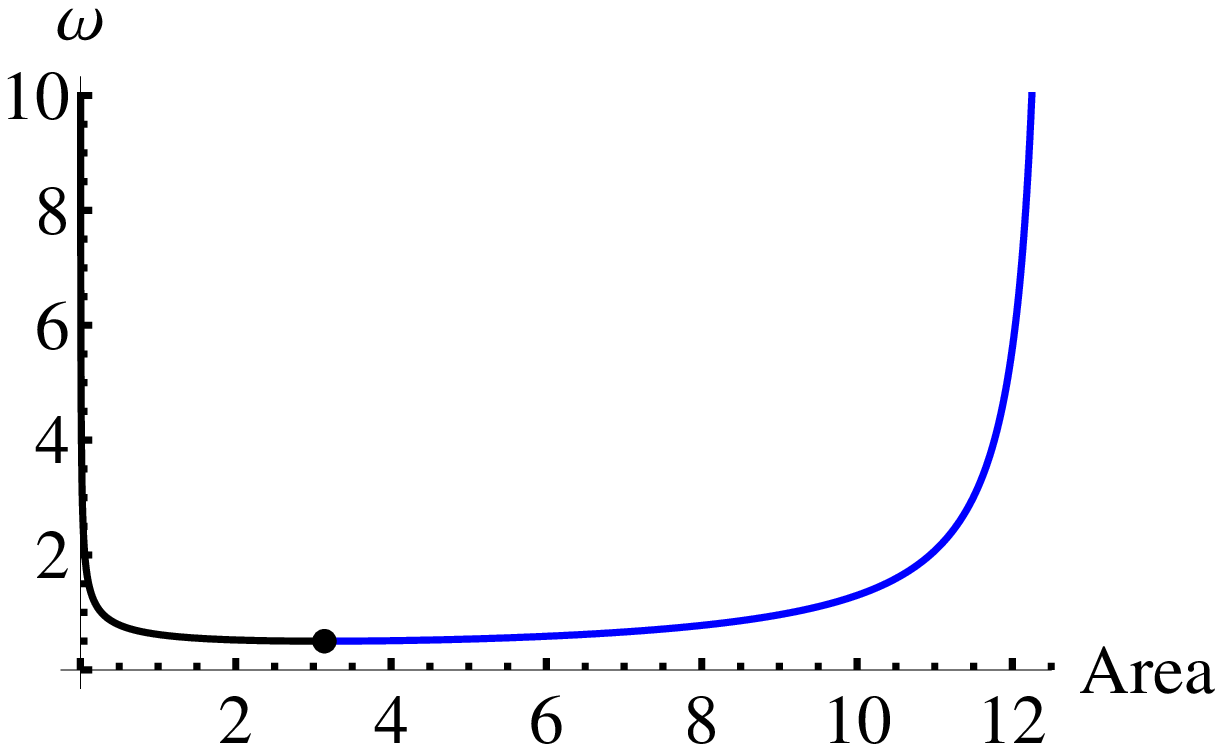}
\caption{The position of two horizons via $\omega$(left) and capture-cross section for different $\omega\,$ values(right).}
\label{fig:hori}
\end{figure}

Horizon size affects capture-cross section. It is the area that the particle cannot escape from the gravity of black hole, and proportional to square of horizon size. The Schwarzschild black hole has only one capture-cross section for particles coming from outside of horizon. The KS black hole gives two capture-cross section from inner and outer horizon dependent on the constant $\omega\,$. Like horizons behavior, outer capture-cross section is closed to Schwarzschild's capture-cross section, and area of inner one becomes zero. It is shown in Fig.(\ref{fig:hori}). The horizon size is always smaller than that of Schwarzschild case, because of effect from modifications. The inner horizon naturally makes particles inside of outer horizon not touch the singularity.

Geometry can be tested by the probe particle. The geodesics are obtained by solving the equation of motion. The  geodesics of probe particle motions can be classified by effective potential.  The formal analysis are used to find to the equations\,\cite{Gwak}. The geodesic equations are derived from the action written,
\begin{eqnarray}
\label{geolag1}
\mathcal{L}=\frac{1}{2} g_{\mu \nu} \dot{x}^\mu\dot{x}^\nu&=&\frac{1}{2}\left(-f(r)\dot{t}^2+\frac{\dot{r}^2}{f(r)}+r^2\dot{\theta}^2+r^2 \sin^2\theta \dot{\phi}^2\right)=\frac{1}{2}\beta\,,
\end{eqnarray}
where the dot notation is $\partial/\partial\lambda$. The conserved quantities are defined to $E$ and $L$ for $t$ and $\phi$ equations of motions. It gives a expression for $\dot{t}$ and $\dot{\phi}$ which is inserted to eq.(\ref{geolag1}). The Lagrangian is rewritten, 
\begin{eqnarray}
\mathcal{L}=\frac{1}{2}\left(-\frac{E^2}{f(r)}+\frac{\dot{r}^2}{f(r)}+r^2\dot{\theta}^2+\frac{L^2}{r^2\sin^2\theta}\right)=\frac{1}{2}\beta\,,
\label{action1}
\end{eqnarray}
in which the parameter, $\beta$, can be 0 for null and -1 for timelike geodesics. The coordinate $\theta$ is fixed at $\theta=\frac{1}{2}\pi$ using spherical symmetry without loss of generality. The equation of motions are used to obtain above result. The equation of motion for radial coordinate is given,
\begin{eqnarray}
\frac{\ddot{r}}{f(r)}-\frac{\dot{r}^2 f'(r)}{f(r)^2}-\frac{f'(r)\left(E^2-\dot{r}^2\right)}{2f(r)^2}-\frac{L^2}{r^3}=0\,,
\end{eqnarray}
in which $\,'=\partial/\partial r\,$. The eq.(\ref{action1}) can be rewritten as,
\begin{equation}
\label{effectiveequation}
\frac{1}{2}\dot{r}^2+\frac{1}{2}f(r)\left(\frac{L^2}{r^2}-\beta\right)=\frac{1}{2}E^2\,,
\end{equation}
where $V_{KS}=\frac{1}{2}f(r)\left(\frac{L^2}{r^2}-\beta\right)\,$ is called effective potential. In large $\omega$ or $r$ limit, the function $f(r)$ becomes Schwarzschild case. In this case, the effective potential becomes $V_{Schwarzschild}=\frac{1}{2}(1-\frac{1}{r})\left(\frac{L^2}{r^2}-\beta\right)\,$. The full set of geodesic equations is derived using equation of motion and eq.(\ref{effectiveequation}). The equations are written,
\begin{eqnarray}
\label{eq:geo}
&&\frac{dt}{dr}=\frac{\dot{t}}{\dot{r}}=\frac{E}{f(r)\sqrt{E^2-f(r)\left(\frac{L^2}{r^2}-\beta\right)}}\,,\label{eqgeo}\\
&&\frac{d\phi}{dr}=\frac{\dot{\phi}}{\dot{r}}=\frac{L}{r^2\sqrt{E^2-f(r)\left(\frac{L^2}{r^2}-\beta\right)}}\,,\nonumber\\
&&\frac{dt}{d\phi}=\frac{\dot{t}}{\dot{\phi}}=\frac{E r^2}{L f(r)}\,.\nonumber
\end{eqnarray}
The geodesic orbit is obtained from modified eq.(\ref{eq:geo}),
\begin{eqnarray}
\label{eq:geo1}
\frac{d^2 r}{d\phi^2}=\frac{2r^3}{L^2}\left[E^2-2f(r)\left(\frac{L^2}{r^2}-\beta\right)\right]+\frac{r^4}{L^2}\left[-f'(r)\left(\frac{L^2}{r^2}-\beta\right)+2f(r)\frac{L^2}{r^3}\right]\,.
\end{eqnarray}

Effective potentials of both KS and Schwarzschild black hole are given,
\begin{eqnarray}
\label{eq:vkschw1}
&&V_{KS}=\frac{1}{2}(4\omega L^2-\beta)+\frac{L^2}{2r^2}-2\omega \beta r^2 -2\omega L^2 \sqrt{1+\frac{1}{2\omega r^3}}+2 \beta \omega r^2 \sqrt{1+\frac{1}{2\omega r^3}}\,,\\
&&V_{Schwarzschild}=\frac{1}{2}\left(1-\frac{1}{r}\right)\left(\frac{L^2}{r^2}-\beta\right)=-\frac{\beta}{2}+\frac{\beta}{2r}+\frac{L^2}{2r^2}-\frac{L^2}{2r^3}\,\nonumber.
\end{eqnarray}
In $\omega\rightarrow\infty$ limit, the effective potential of KS case become to that of Schwarzschild case in eq.(\ref{eq:vkschw1})\,.
\begin{figure}[h]
\centering
\includegraphics[scale=0.52,keepaspectratio]{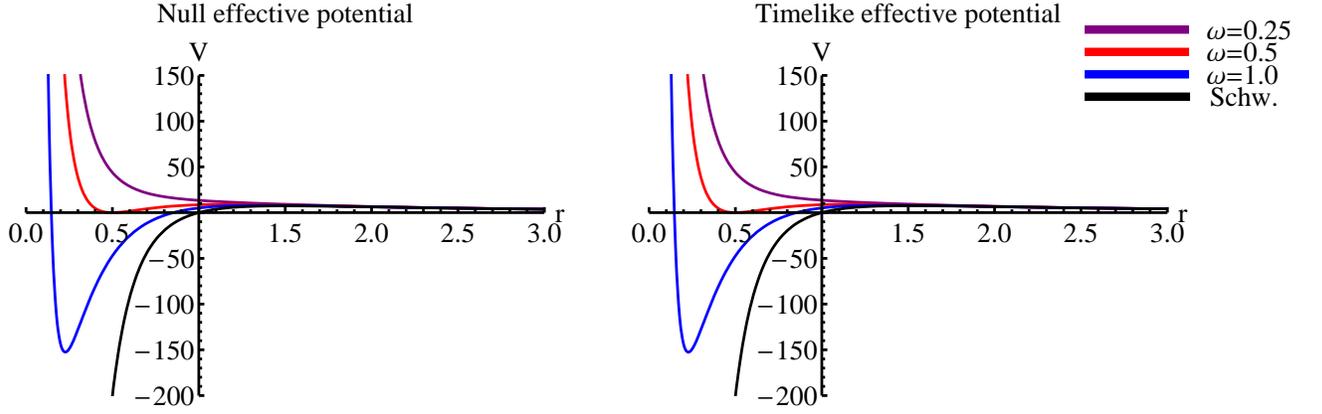}
\caption{The effective potential for $L$=10 is plotted for different values of $\omega$ and for Schwarzschild black hole.}
\label{fig:nteffec}
\end{figure}
The structures of effective potential is given in Fig.\ref{fig:nteffec} for different $\omega\,$ values. In HL-theory, the effective  potentials fall slower than that of  Schwarzschild case. Because of repulsive force from singularity, the particle cannot fall to center. Increasing value of $\omega$ makes this black object being close to Schwarzschild black hole. The trend is also shown by effective potential getting close to Schwarzschild case for large $\omega\,$ values. The interesting point is the positive value of effective potential around $r$=0, repulsive singularity which exists independence of $\omega\,$. Therefore, repulsive force acts to the null or timelike objects which are close to singular region, making them scattered or trapped. This phenomena is similar to the repulsive singularity of Reissner-Nordstr©ªm black hole, even though KS black hole has no electric charges. As an examples, naked singularity for, $\omega$=0.25, protects its singularity by repulsive effective potential. Effective potential is monotonically decreased from the singularity at $r=0$ to $r=\infty$. When this solution has the one horizon at $\omega=$0.5, the minimum point is $V_{KS}=0$ positioned at $r_-$, and the maximum point is located at $r_+$ in the shape of effective potential. In Fig.(\ref{fig:nteffec}), there exists minimum point of $V_{KS}$ if $\omega > 0.385$ for null case and $\omega > 0.383$ for timelike case. The generic value of $\omega$ is related to angular momentum $L$. The KS effective potential is matched to Schwarzschild case in $\omega\rightarrow\infty$ limit except for positive cusp at $r=0$. 

The effective potentials are expanded at $r=0$ for KS black hole and the other two black holes as,
\begin{eqnarray}
&&V_{KS}=\frac{L^2}{r^2}-\frac{L^2\sqrt{2\omega}}{r^{3/2}}+(4L^2\omega-\beta)+O(r)\,,\\
&&V_{Schwarzschild}=-\frac{\beta}{2}+\frac{\beta}{2r}+\frac{L^2}{2r^2}-\frac{L^2}{2r^3}\,,\nonumber\\
&&V_{RN}=\frac{1}{2}\left(1-\frac{2M}{r}+\frac{e^2}{r^2}\right)\left(\frac{L^2}{r^2}-\beta\right)=\frac{e^2L^2}{2r^4}-\frac{ML^2}{r^3}+\frac{L^2-e^2\beta}{2r^2}+\frac{M\beta}{r}-\frac{\beta}{2}\,.\nonumber
\label{eq:veff3}
\end{eqnarray}
The most dominant terms depend on angular momentum, $L$. The term is always positive around singularity at $r=0$ for KS and RN Black hole case. In fact, the singularity of KS black hole is repulsive, and there are two horizons like RN black hole. The power of  dominant terms for KS black hole is smaller than that for RN black hole. Around center, the repulsive force for acting to geodesics of KS case is weaker than that of RN case. The power of repulsive force for KS black hole, $F=-\frac{\partial V_{eff}}{\partial r}$, around singularities is $r^{-3}$, and that of RN black hole is $r^{-5}$. Different from charge dependent RN black hole, the singularity of KS black hole is geometrical property. The attraction force is dominant at Schwarzschild and 2nd dominant at RN and KS case. The attractive force of Schwarzschild case is stronger than that of KS and RN case. Also, it can be shown at Fig.(\ref{fig:nteffec}). The feature is the power of the attraction force of KS. Gravitational attraction is weaker than that of Schwarzschild case. It gives interesting points that the gravitation becomes weaker at around and in the outer horizon than that of Schwarzschild case, and the plus sign of cusp-like shape of potential in KS black hole makes empty space around $r$=0 region in which matter cannot stay, but it feels repulsive force. It is very different situation from the minus sign cusp-like shape of potential in Schwarzschild contains huge mass density around center. 
\begin{figure}[h]
\centering
\includegraphics[scale=0.5,keepaspectratio]{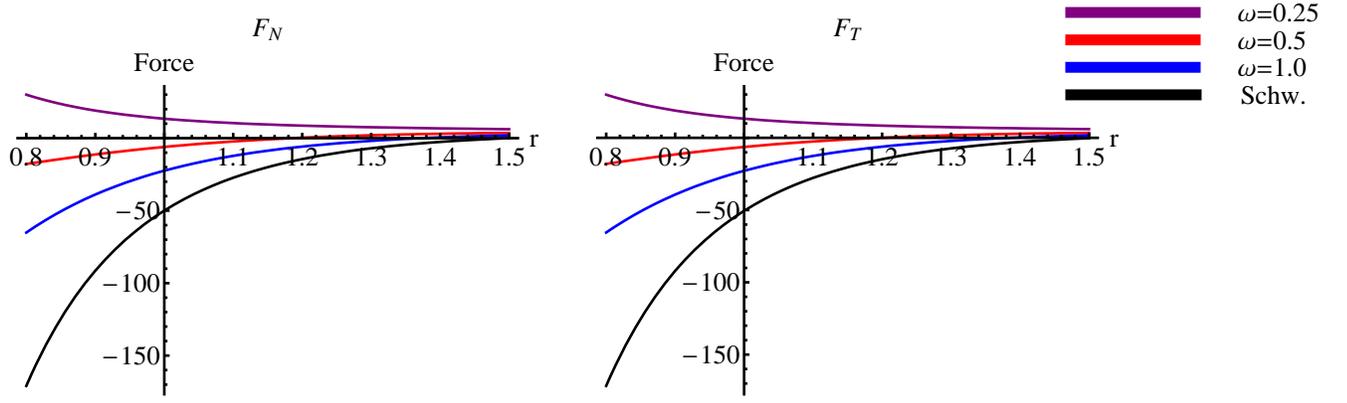}
\caption{There are effective force plots for $L=$10 around Schwarzschild horizon $r=1$. The gravitational forces are $F_{N}$ for null case and $F_{N}$ for timelike case. Those have $\omega$-dependency.}
\label{fig:frcpl}
\end{figure}
The gravitational force $F$ around horizon is in Fig.(\ref{fig:frcpl}). The gravitation forces of KS black hole are similar to that of Schwarzschild case at far from horizon. The probe around horizon is monotonically repulsed in small $\omega$. Increasing $\omega$ enforces attraction around outer horizon. After $\omega \ll\,$0.5, attractive force acts like Schwarzschild black hole. In the limit of $\omega\rightarrow\infty$, the size of the force is closed to Schwarzschild case. The KS black hole gravity is always weaker than Schwarzschild around horizon in Fig.(\ref{fig:frcpl}). In this context,  weaker gravity causes the density or velocity of the matters which is accumulated around horizon. KS black hole makes matter density smaller and broader than Schwarzschild black hole. Because the centrifugal force is weaker, the velocity of the gas clouds circulating around the black hole is slower than that of Schwarzschild solution.

\begin{figure}[h]
\centering
\includegraphics[scale=0.3,keepaspectratio]{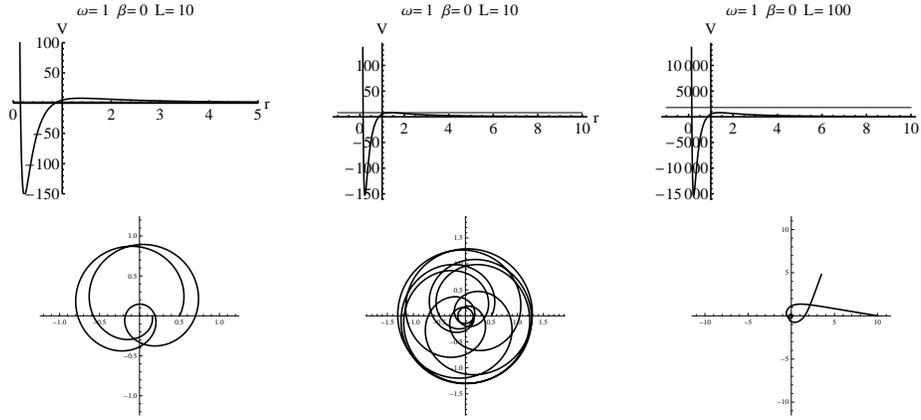}
\caption{Null geodesic orbits for different energy and $\omega$=1 which is two horizons. There are closed precession, the other precession orbit, and scattering.}
\label{fig:np1s}
\end{figure}

\begin{figure}[h]
\centering
\includegraphics[scale=0.3,keepaspectratio]{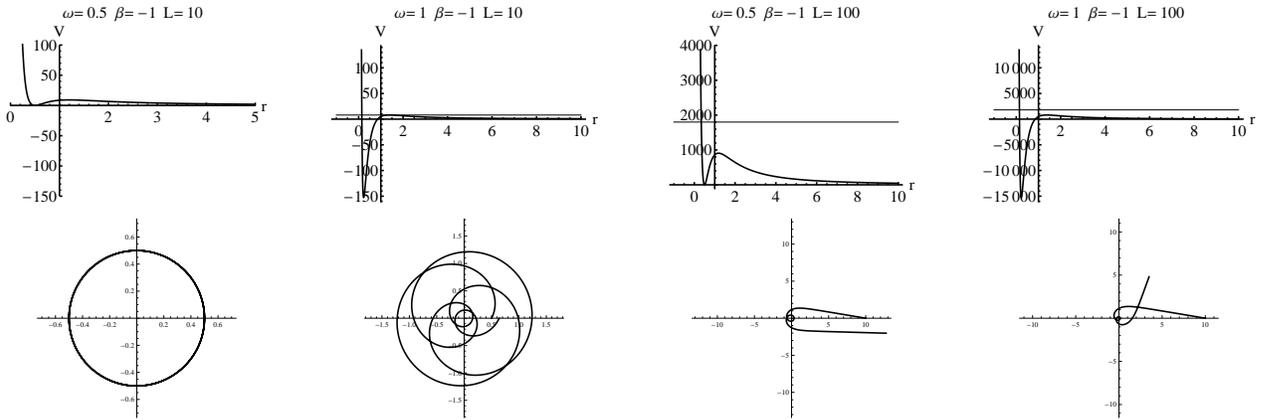}
\caption{Timelike geodesic orbits for different energy and $\omega$=0.5, 1.0. There are stable circular orbits, closed precession orbits, and scattering orbits.}
\label{fig:tp1s}
\end{figure}

We now study geodesic orbits. Light or matter can be scattered or has a closed orbit in and around its horizons. It makes very different universe generated from Schwarzschild's, since the energy emitted from KS black hole is larger than expectation from Schwarzschild case. In this sense, The effective potential only gives a part of information about particle orbits. The full information is included on geodesic equation, eq.(\ref{eq:geo}).
\begin{figure}[h]
\centering
\includegraphics[scale=0.3,keepaspectratio]{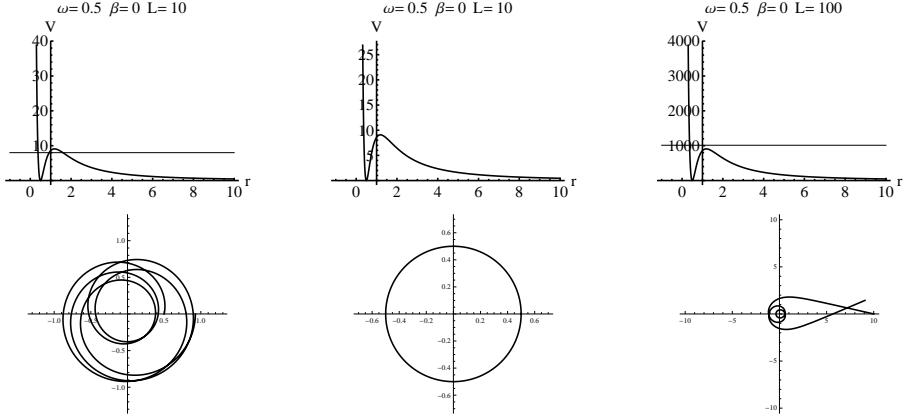}
\caption{Null geodesic orbits for different energy and $\omega$=0.5 which is only one horizon. There are closed precession, stable circular orbit, and scattering.}
\label{fig:np05s}
\end{figure}
The new orbits are classified by $\omega$ constants. There are null orbits of one horizon case, $\omega=0.5$, represented in Fig.\ref{fig:np05s}. There are closed orbits. The first figure represents a precession with orbit, period $4\pi\,$, and the 2nd one is a stable circular orbit, which is not observed in Schwarzschild black hole. Also, it has unstable circular orbit at maximum point of effective potential that is not included in the figures. As expected, the light cannot arrive at $r=0$, but  scattered or trapped in potential well, due to the  potential wall. This property implies that it has very different interior physical structure from that of Schwarzschild case. There are no null geodesics falling to center. For the constant $\omega=1$, two horizon, the precession, unstable circular, and scattering orbits are possible in null geodesics as shown in Fig.\ref{fig:np1s}. The stable bounded precession orbits and scattering orbits are possible, but the stable circular motion does not exist, because the minimum point of effective potential is positioned at the point of minus sign. Only unstable circular motion can exist in this parameter set. The other feature is the scattering orbits. The effective potential has repulsive singularity which do not permit the particles to enter singular region. The particles cannot fall to center. This property is not observed in Schwarzschild black hole. The scattering is related to gravitational lensing effects\,\cite{bib5}. The spacetime structure might be observable to asymptotic observer who measures the bending of light. The trend of timelike  orbit behaviors is similar to that of null case in Fig.\ref{fig:np05s}, \ref{fig:np1s} and shown in \ref{fig:tp1s}. Massive particles also cannot touch the singularity. 

\begin{figure}[h]
\centering
\includegraphics[scale=0.50,keepaspectratio]{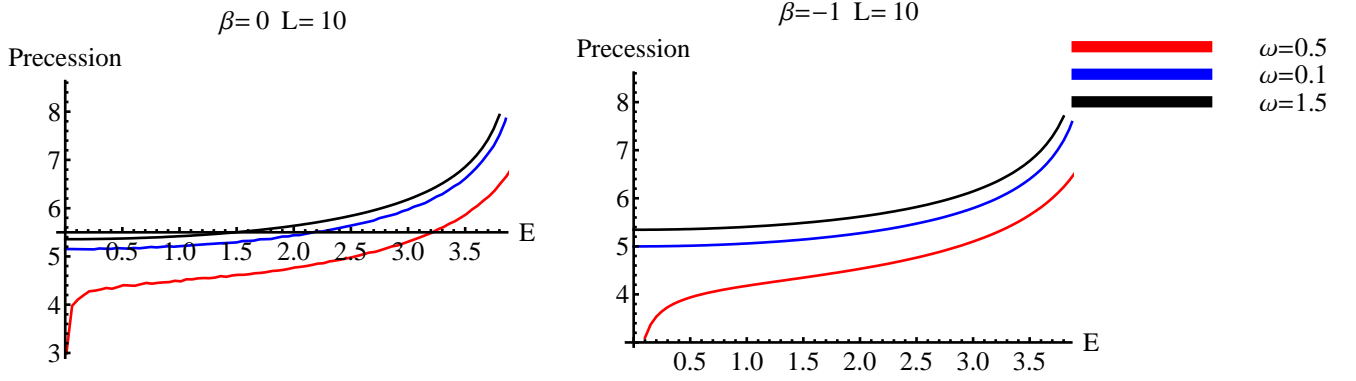}
\caption{The orbits which are dependent on the $\omega$ are classified by precession rate. Left is plotted for null geodesics and right is plotted for timelike geodesics. Each lines is plotted under changing energy of particles in fixed angular momentum.} 
\label{fig:pres}
\end{figure}

The precession rates for these orbit is affected by coupling constants. Each geodesic orbits is dependent on $\omega$ and their energy. Their behaviors can be  classified using precession rates of each orbits. The different $\omega$ constants represent different spacetimes which have different couplings with higher curvature terms. The Fig.\ref{fig:pres} is plotted for six cases classified under timelike and null case for different $\omega\,$. The precession orbits which cannot be observed in Schwarzschild black hole are numerically evaluated in Fig.\ref{fig:pres}. Although the KS black hole is asymptotically Schwarzschild solution, the geodesic orbits around horizon behave differently from that of Schwarzschild case. The precession rates is increased in the large values of $\omega$. It is the open orbits in Schwarzschild spacetime that the precession rate is infinity in the precession point of view. For large $\omega$, the precession rate of KS black hole tends to also match to Schwarzschild case by larger rate. Especially, one horizon case, $\omega=$0.5, has a stable circular orbit in both null and timelike geodesics, so the precession rate of $E=0$ starts at zero. Since the other case do not have such orbit, they do not need to start at zero. 

\begin{figure}[h]
\centering
\includegraphics[scale=0.65,keepaspectratio]{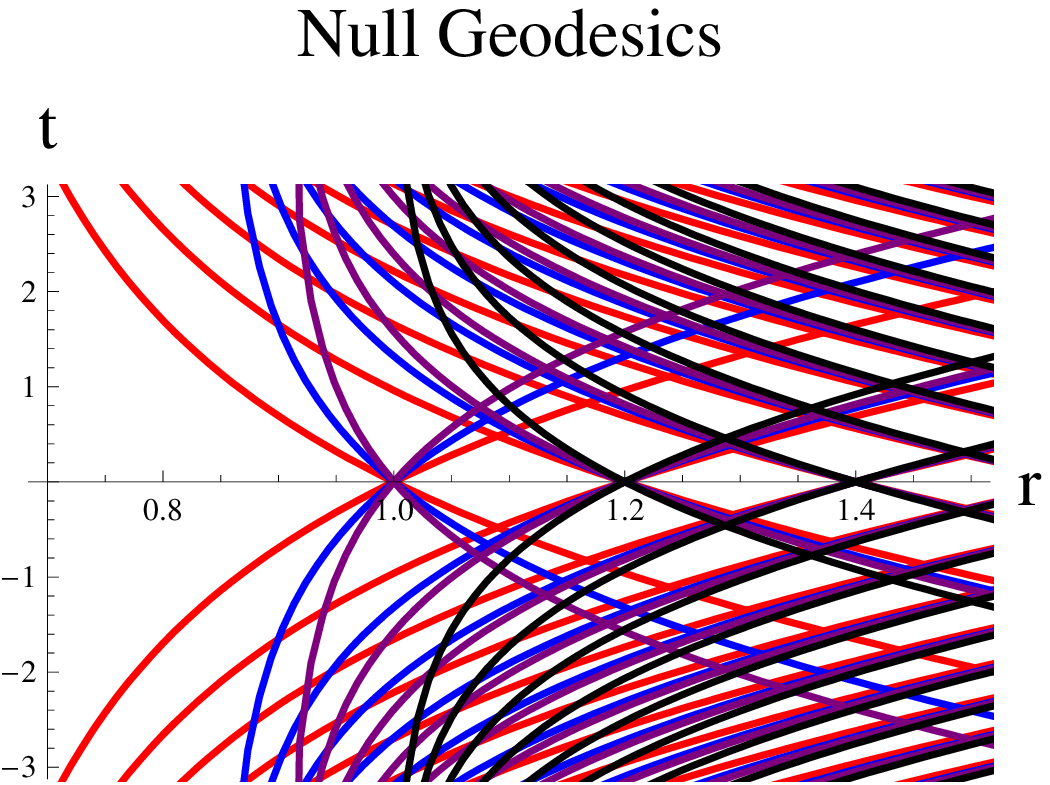}
\includegraphics[scale=0.65,keepaspectratio]{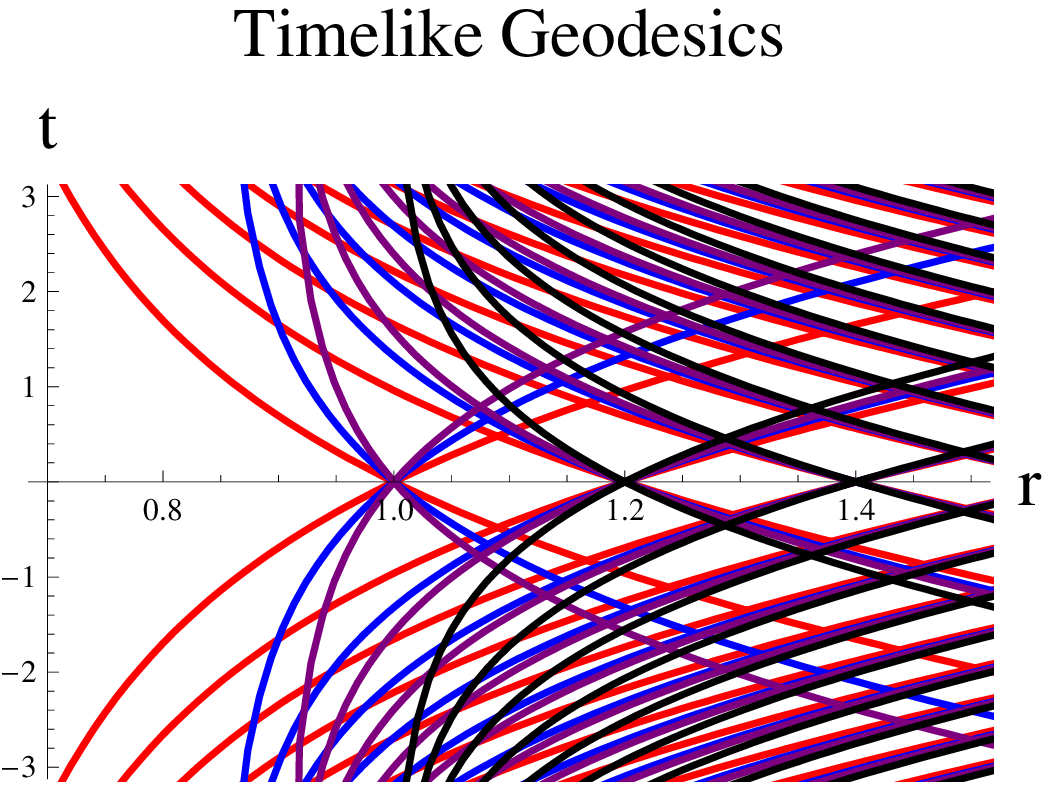}
\caption{The classified radial in-going and out-going null(left) and timelike(right) geodesics for values of $\omega\,$=0.5(red), $\omega\,$=1.0(blue), $\omega\,$=1.5(purple), and Schwarzschild(black).}
\label{fig:trgeo}
\end{figure}

The time trajectories are dependent on the constant, $\omega\,$. However, the contribution of higher curvatures deforms geometry and gives different time structure around outer horizon. The in-going and out-going null and timelike trajectories are written from time-radial motion geodesics eq.(\ref{eq:geo}). The in-going and out-going radial null and timelike geodesics are given Fig.\ref{fig:trgeo}. Increasing $\omega$ makes the KS spacetime more curved while it becomes the geometry of Schwarzschild. It means that the KS spacetime is flatter than Schwarzschild's. Precisely, the geodesics started at same distant from center are more curved at larger $\omega$. The location of outer horizon which is dependent on $\omega$ is close to center for small $\omega$. It makes geodesics come into closer to center.At a fixed point, the spacetime becomes flatter at larger value of $\omega$. The geodesic effect contributes to weaker gravitation force around and in horizons.

\section{Conclusion}
The geodesic properties of KS black hole is studied. The KS black hole solution is flatter compared with Schwarzschild black hole in and around horizon. The probe far from center of KS black hole behaves similar to Schwarzschild case. In the close region to center, the gravitation becomes weaker compared with Schwarzschild's by the effect of flatter geometry. In the effective potential, KS solution has repulsive nature at singularity like RN black hole. KS effective potential matches to Schwarzschild case in $\omega=\infty$. Compared with $V_{RN}$, the repulsive singularity of KS black hole is weaker than that of RN black hole. Around horizon, the gravitation becomes weaker than Schwarzschild solution. The attraction of Schwarzschild black hole is stronger than that of KS solution. The weaker KS gravitation around horizons will attract small mass. The formation and mass of matter around horizon is changed under this effects. There are no particle fall to center like Schwarzschild black hole, but particles are scattered to infinity or trapped periodical motions. This orbits, scattering and periodical motions do not exist at Schwarzschild black hole. New orbits are a result of competition between attraction and repulsive force around horizons. These orbits are numerically obtained and classified under $\omega$ for null and timelike case. The periodical motions are consist of a stable circular orbit at $\omega$=0.5 and elliptic orbits with precession. The precession orbits are classified to six cases under $\omega$, timelike, and null. Since repulsive potential is at $r=0$, all of null and timelike geodesics suffer the scattering motions even if high energy. They cannot see the singularity. It implies that the geometry and interior of black hole are different far from Schwarzschild solution. Since gravitational potential of KS black hole has repulsive singularity, it cannot make dense mass staying in its center unlike Schwarzschild's, so the total mass of trapped matter in the black hole is supposed to be smaller than expected. In addition to classification of orbits, the geodesic time structures are classified using in and out going null and timelike geodesics. The geodesic lines are close to Schwarzschild in larger $\omega$. The outer horizon closes to center in small $\omega$, so time geodesics in small $\omega$ can reach than that in large $\omega$.\\

{\bf{Acknowledgments}}

We would like to thank Gungwon Kang, Seoktae Koh, Wonwoo Lee, and Changyong Park for their kind comments. This work was supported by the Korea Science and Engineering Foundation (KOSEF) grant funded by the Korea government(MEST) through the Center for Quantum Spacetime(CQUeST) of Sogang University with grant number R11-2005-021.\\

\end{document}